\documentclass[journal]{IEEEtran}
\ifCLASSINFOpdf
\usepackage[pdftex]{graphicx}
\graphicspath{figures/}
\DeclareGraphicsExtensions{.pdf,.jpeg,.png}
\else
\usepackage[dvips]{graphicx}
\graphicspath{{figures/}}
\DeclareGraphicsExtensions{.eps}
\fi
\usepackage[cmex10]{amsmath}
\usepackage{bm}
\usepackage{subfigure}
\usepackage{hyperref}
\usepackage{mathtools}
\usepackage{cite}
\DeclareMathOperator{\sech}{sech}
\usepackage{amsfonts}
\usepackage{url}
\usepackage[table,xcdraw]{xcolor}

\begin{document}

\title{Emulation of Astrocyte Induced Neural Phase Synchrony in  Spin-Orbit Torque Oscillator Neurons}

\author{Umang~Garg, Kezhou~Yang, 
and~Abhronil~Sengupta
\thanks{The authors are with the School
of Electrical Engineering and Computer Science, Department of Materials Science and Engineering, The Pennsylvania State University, University Park,
PA 16802, USA. U. Garg is also affiliated with Birla Institute of Technology and Science, Pilani, Rajasthan 333031, India. E-mail: sengupta@psu.edu.
}}
\maketitle
\begin{abstract}
Astrocytes play a central role in inducing concerted phase synchronized neural-wave patterns inside the brain. In this article, we demonstrate that injected radio-frequency signal in underlying heavy metal layer of spin-orbit torque oscillator neurons mimic the neuron phase synchronization effect realized by glial cells. Potential application of such phase coupling effects is illustrated in the context of a temporal ``binding problem". We also present the design of a coupled neuron-synapse-astrocyte network enabled by compact neuromimetic devices by combining the concepts of local spike-timing dependent plasticity and astrocyte induced neural phase synchrony.
\end{abstract}


\maketitle
%

\section{\label{sec:level1}Introduction}

Neuromorphic engineering is emerging to be a disruptive computing paradigm in recent times driven by the unparalleled efficiency of the brain at solving cognitive tasks. Brain-inspired computing attempts to emulate various aspects of the brain's processing capability ranging from synaptic plasticity mechanisms, neural spiking behavior to in-situ memory storage in the underlying hardware substrate and architecture. The work presented in this article is guided by the observation that current neuromorphic computing architectures have mainly focused on emulation of bio-plausible computational models for neuron and synapse – but have not focused on other computational units of the biological brain that might contribute to cognition.


Over the past few years, there has been increasing evidence that glial cells, and in particular, astrocytes play an important role in multitude of brain functions \cite{allam2012computational}. It is estimated that glia form approximately 50\% of the human brain cells \cite{ glial_cell_information_routing}
and participate by modulating the neuronal firing behaviour, though unable to discharge electrical impulses of their own. Indeed, these glial-cells work in coordination with neural assemblies, to enable information processing in the human brain and performing incisive operations. Astrocytes hold the recipe to potentiate or suppress neurotransmitter activity within  networks and are responsible for phenomenon like synchronous network firing \cite{Nature_2011, A.P_citation2} and self-repair mechanisms \cite{Astrocyte_self-repair_bio2,rastogi2020self}. It is therefore increasingly important to capture the dynamics of such ensembles, a step towards realizing more sophisticated neuromimetic machines and ultimately enabling cognitive electronics.

Recently, there has been extensive literature reporting astrocyte computational models and their impact on synaptic learning \cite{compmodels,de2012computational}. Continuing these fundamental investigations to decode neuro-glia interaction, there have been recent neuromorphic implementations of astrocyte functionality in analog and digital Complementary Metal Oxide Semiconductor (CMOS) hardware \cite{Glia_in_Self_repairing_SNN,glial_cell_information_routing, alice15, KARIMI_elsevier,  Ranjbar_springer2,Bernabe_AFM}. 
For instance, analog CMOS circuits capturing the neural-glial transmitter behaviour have been demonstrated \cite{alice11, alice13, alice15, alice16}. There is also  increasing interest in low-complexity FPGA implementation of the astrocyte computation models \cite{KARIMI_elsevier, NAZARI_elsevier, Ranjbar_springer1, Ranjbar_springer2,Bernabe_AFM}.
However, the primary focus has been on a brain-emulation perspective, i.e. implementing astrocyte computational models with high degree of detail in the underlying hardware.

On the other hand, recent advances in emerging post-CMOS technologies like phase change materials, resistive memories, ferromagentic and ferroelectric materials \cite{jackson2013nanoscale,kuzum2011nanoelectronic,jo2010nanoscale,ramakrishnan2011floating,sengupta2017encoding,saha2021intrinsic}, among others have resulted in the development of electronic device structures that can reproduce various biomimetic characteristics at low operating voltages through their intrinsic physics.
However, while there has been extensive work on exploring post-CMOS technologies for mimicking bio-realistic computations due to the prospects of low-power and compact hardware design, they have been only studied from standalone neuron/synapse perspective. Emulation of the neuron-astrocyte crosstalk using bio-mimetic devices has largely been neglected, and no such literature exists hitherto, to the best of our knowledge. This work is therefore an effort to bridge this gap and, specifically, elucidates the emulation of transient synchronous activity resulting from neural-glial interactions by utilizing spin-orbit torque induced phase synchronization of spintronic oscillator neurons. It is worth mentioning here that we abstract the neuron functionality as a non-linear oscillator, in agreement with prior neuroscience and computational models \cite{Neuron_as_oscillator3}. Emulation of astrocyte induced neural phase synchrony through the intrinsic physics of spintronic devices will be critical to enable the next generation of resource constrained cognitive intelligence platforms like robotic locomotion \cite{polykretis2020astrocyte}. This work also presents an important addition to the wide variety of next-generation computational paradigms 
like associative computing, vowel-recognition, physical reservoir computing among others \cite{Vowel_recognition_STO,KaushikRoy_STOsync,Physical_Reservoir_computing_STO, Binding_Events_STO,riou2019temporal,torrejon2017neuromorphic}, being implemented using spin-torque oscillator devices.
\section{\label{sec:Network Architecture} Neuroscience Background}
The human brain houses multiple-independent local neuronal groups which perform dedicated computations in relevance to their assigned tasks.
Besides this general uncorrelated activity of neurons, multiple neural spiking data recordings reveal that the independent signals from these neural assemblies frequently coalesce in time to generate a synchronous output \cite{Nature_2011, sync_firing1}. 
Multiple reports on the cause of such patterns now provide compelling evidence that astrocytes are the agents of this phenomenon \cite{A.P_citation1, A.P_citation2}. Astrocytes modulate the  concentration of neurotransmitters like glutamate inside the synaptic clefts in response to its internal Calcium ($Ca^{2+}$) oscillations \cite{astro_Ca_NT_intrxn_1,astro_Ca_NT_intrxn_2}. A single astrocyte spans tens of thousands of synapses, where units called microdomains (concentrated $Ca^{2+}$ stores within the astrocyte) monitor the activity for a group of neurons and perform subsequent chemical actions \cite{Astrocyte_microdomains,astrocyte_microdomains2}.
The astrocyte-derived glutamate binds to extrasynaptic NMDAR (N-methyl-D-aspartate) receptor channels, and induce Slow-inward Currents (SIC) in the post-synaptic membrane.
SICs are attributed to triggering a simultaneous response in different synapses with high timing precision, and its large amplitude and slow-decay rate provide an increased timescale for the correlated activity \cite{A.P_citation1, A.P_citation2}. 
The astrocytic units influencing synapses, can act both independently or in coordination enabling long-distance indirect signalling among independent neuronal groups. Furthermore, an increased intensity of synaptic activity can trigger multiple astrocytes to share their information through their gap-junctions and elicit coherent behaviours among different uncorrelated neuronal networks.  We in this paper do not discriminate among the two signalling processes. Thus the two astrocytes shown in Fig. 1(a) for different sub-networks can also imply microdomains within a single astrocyte. These units control the synchronization signal to networks A and B.
Fig. 1(a) captures the biological perspective of such a system which controls the neural synchronization among neurons present in these different sub-networks. Sub-networks A and B each consist of three different neurons, which in-turn generate oscillatory outputs.
The temporal profiles, shown in Fig. 1(b), depict the neuron outputs before and after synchronization is initiated by Astrocyte 1 in the network A. Interested readers are referred to Ref. \cite{A.P_citation2} for details on the astrocyte computational models. It is worth mentioning here that unlike CMOS implementations that are able to implement computational models with a high degree of detail, emerging device based implementations usually focus on mimicking key aspects of the neurosynaptic functionality necessary from computing perspective since the exact behavior is governed by the intrinsic device physics. In this work, we primarily consider emulating the neural phase synchrony effect of astrocytes and evaluate it in the context of a temporal information binding application.

\begin{figure}[t]
\center
  \subfigure[]{\includegraphics[width=0.95\columnwidth]{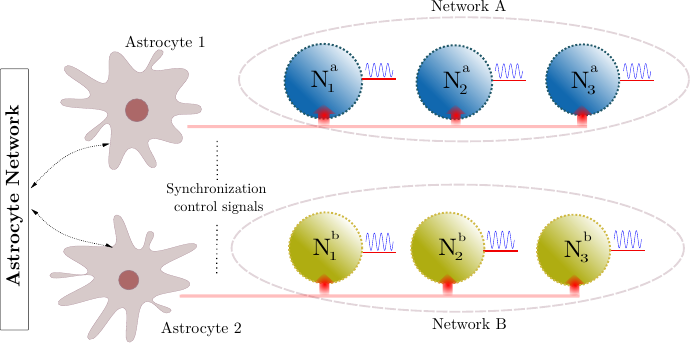}\label{fig:top_level_diagram_part1}}\quad 
    \subfigure[]{\includegraphics[width=0.95\columnwidth]{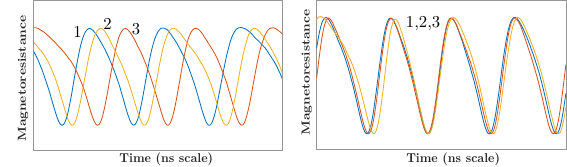}\label{fig:top_level_diagram_part2}}
\caption{(a) Top-level network depicting the synchronization control by astrocytic injection. Astrocytes share information among their glial network. (b) The curves show the synchronized and unsynchronized outputs of Neurons 1-3 in Network A depending on the astrocyte input.}
\end{figure}
\section{Astrocytic Synchronization Emulation }
\subsection{Device Basics}
In this work, we utilize Magnetic Tunnel Junctions (MTJs) \cite{MTJ_des} as the core hardware primitive to mimic neural oscillations. 
The MTJ consists of two ferromagnetic layers (pinned layer and free layer) with a spacer oxide layer in between. The direction of magnetization of the pinned layer (PL) is fixed, while that of the free layer (FL) can be manipulated by external stimuli (spin current/magnetic field). The MTJ stack exhibits a varying resistance depending on the relative magnetic orientations of the PL and the FL. The extreme resistive states are referred to as the parallel (P) and anti-parallel (AP) states depending on the relative FL magnetization. The magnetization dynamics of the FL can be modeled by Landau-Lifshitz-Gilbert-Slonczewski (LLGS) equation with stochastic thermal noise~\cite{LLG_des}:
\begin{equation} 
\label{eqn:LLG}
\frac{d\hat{m}}{dt}=-\gamma(\hat{m}\times H_{eff})+\alpha(\hat{m}\times \frac{d\hat{m}}{dt})+\frac{1}{qN_s}(\hat{m}\times I_s\times \hat{m})
\end{equation}
In Eq. (\ref{eqn:LLG}), $\hat{m}$ is the unit vector representing the magnetization direction of FL, $H_{eff}$ is the effective magnetic field including thermal noise \cite{scholz2001micromagnetic}, demagnetization field and external magnetic field, $\gamma$ is the gyromagnetic ratio, $\alpha$ is Gilbert's damping ratio, $I_s$ is the spin current, $q$ is the electronic charge, and $N_s=\frac{M_{s}V}{\mu_B}$ is the number of spins in free layer of volume $V$ ($M_{s}$ is saturation magnetization and $\mu_{B}$ is Bohr magneton). If the magnitude of spin current and external magnetic field are chosen appropriately such that the damping due to the effective magnetic field is compensated, a steady procession of the FL magnetization can be obtained. It is worth mentioning here that the intrinsic magnetization dynamics in Eq. (\ref{eqn:LLG}) is used to model the oscillator dynamics. Other variants of oscillatory behavior can be achieved by modified spin device structures \cite{Windmill_oscillator}.

\begin{figure}[b]
\center
\includegraphics[width=0.45\columnwidth]{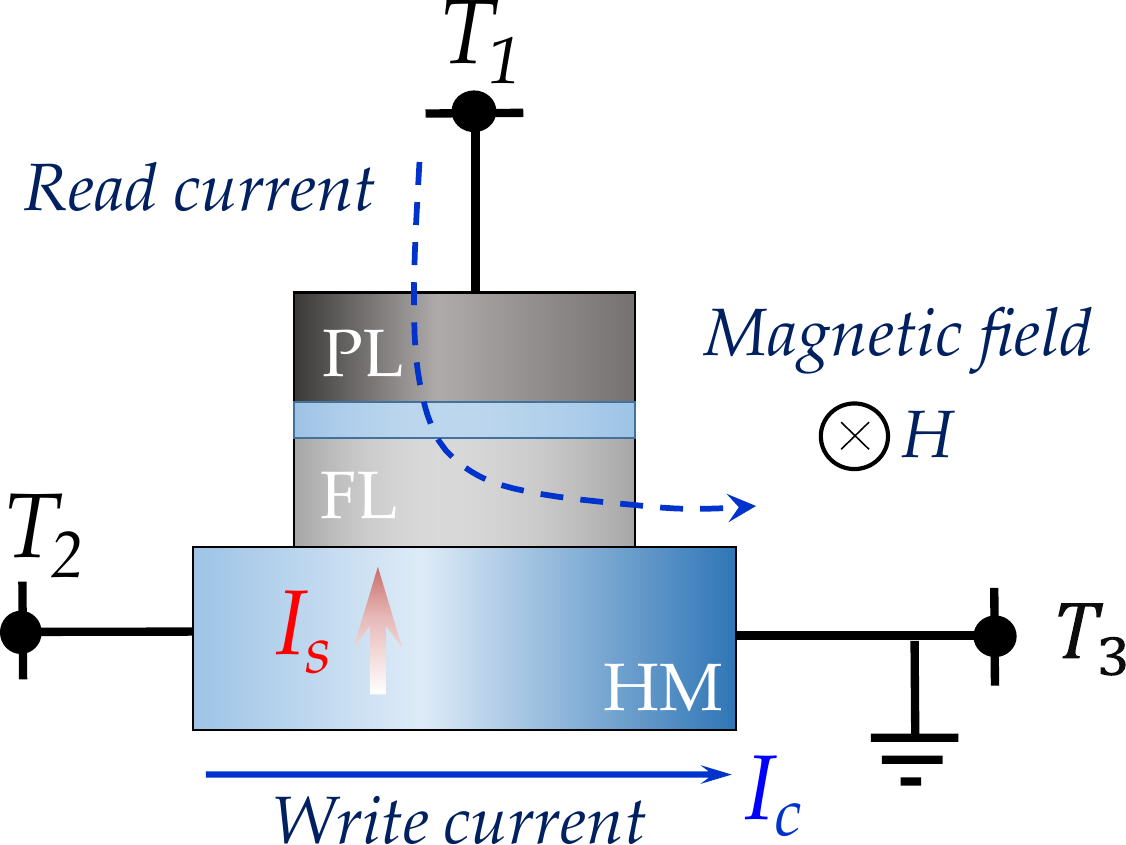}\label{fig:Single_Devices}
\caption{Spin-orbit torque device undergoes oscillation due to applied external magnetic field, $H$, and charge current, $I_c$. Note that the directions of both the magnetic field and magnetic anisotropy are in-plane. }
\end{figure}

In order to achieve decoupled output oscillator readout and astrocyte injection induced phase coupling, we utilize a three terminal device structure, as shown in Fig. 2, in which a nanomagnet with in-plane magnetic anisotropy lies on top of a heavy metal (HM) layer with high spin-orbit coupling. Due to spin-Hall effect \cite{hirsch1999spin}, a transverse spin current is injected into the MTJ FL by charge current, $I_c$, flowing through the HM between terminals T2 and T3. The relation between spin current $I_s$ and charge current $I_c$ is,
\begin{equation} 
\label{eqn:spinangle}
I_s = {\theta_{SH}}\frac{A_{FM}}{A_{HM}}\left(1-\sech \left( \frac{t_{HM}}{\lambda_{sf}} \right) \right)I_c
\end{equation}
where, $A_{FM}$ and $A_{HM}$ are the FM and HM cross-sectional areas respectively, $\theta_{SH}$ is the spin-Hall angle \cite{hirsch1999spin}, $t_{HM}$ is the HM thickness and $\lambda_{sf}$ is the spin-flip length. 
Note that an in-plane magnetic field, $H$, is also applied to achieve sustained oscillation. The MTJ state is read using the current sensed through terminal T1. The device simulation parameters are tabulated in Table I and are based on typical experimental measurements reported in literature \cite{KaushikRoy_STOsync}. However, the conclusions presented in this study are not specific to these parameters. Experimental demonstration of injection locked spin-torque oscillators have been achieved \cite{rippard2013time,rippard2005injection,georges2008coupling,demidov2014synchronization}. It is worth mentioning here that we assume all the devices are magnetically isolated and sufficiently spaced such that dipolar coupling is negligible \cite{yogendra2017coupled}. We also consider that the generated charge current in the HM layer due to FL magnetic precession via the Inverse spin-Hall effect (ISHE) is not dominant enough to impact the phase coupling phenomena. While recent studies have shown that the ISHE modulated current alone, without any amplification, is not sufficient to impact phase locking \cite{elyasi2015synchronization}, such effects can be also overcome by limiting the number of oscillators sharing a common HM substrate. 

\begin{table}[h]
\label{table1}
\center
\centerline{Table I: MTJ Device Simulation Parameters}
\vspace{2mm}
\begin{tabular}{c c}
\hline \hline
\bfseries Parameters & \bfseries Value\\
\hline
Ferromagnet Area, $A_{FM}$ & $40 nm \times 100nm$ \\
HM Thickness, $t_{HM}$ & $3 nm$ \\
Energy Barrier, $E_b$ & 62.76 kT\\
Saturation Magnetization, $M_s$ & $\frac{10^7}{4\pi}$ A/m\\
Spin-Hall Angle, $\theta_{SH}$ & 0.3 \\
Spin-Flip Length, $\lambda_{sf}$ & $1.4nm$ \\ 
Gilbert Damping Factor, $\alpha$ & 0.03 \\
External Magnetic Field, $H$ & 750 Oe \\
TMR Ratio, $TMR$ & 200\% \\
Temperature, $T$ & 300 K\\

\hline \hline
\end{tabular}\\ 
\label{tab:my_label}
\end{table}


\begin{figure}[t]
\center
\includegraphics[width=1\columnwidth]{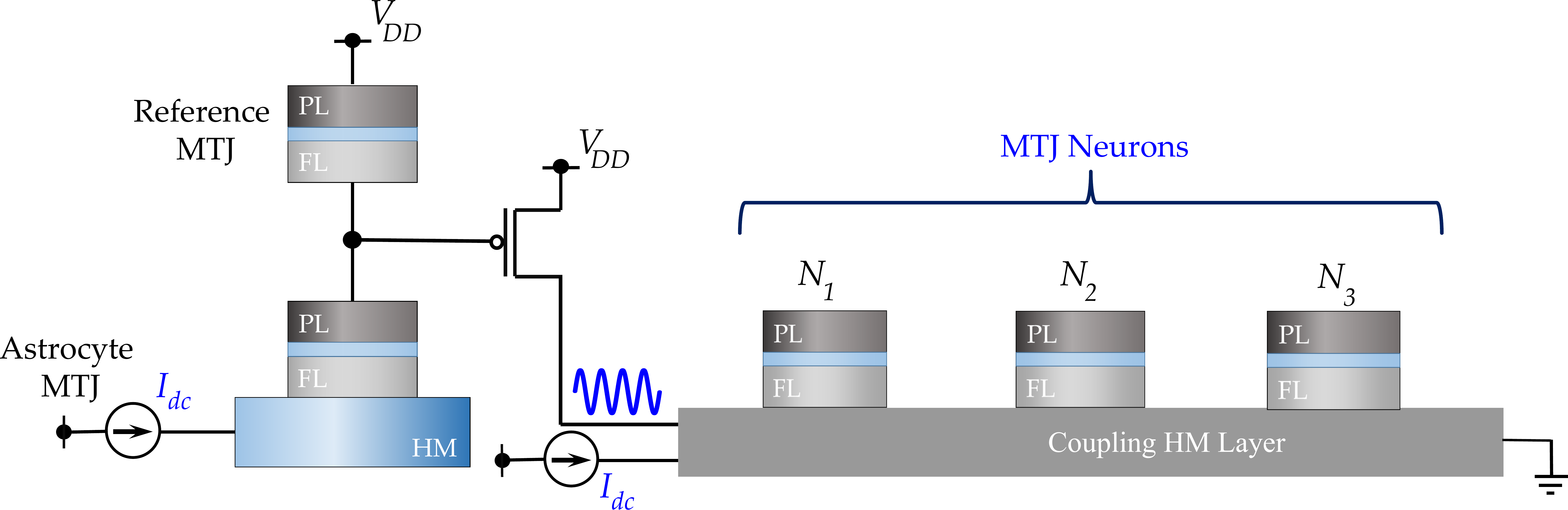}
\caption{\label{fig:Multiple_Devices} Electrical emulation of astrocyte induced neural synchrony is shown where an astrocyte device drives an alternating current through a common HM substrate to phase-lock the MTJ oscillator neurons.}
\end{figure}

\begin{figure*}[bht!]
\center
  \subfigure[]{\includegraphics[width=0.5\columnwidth]{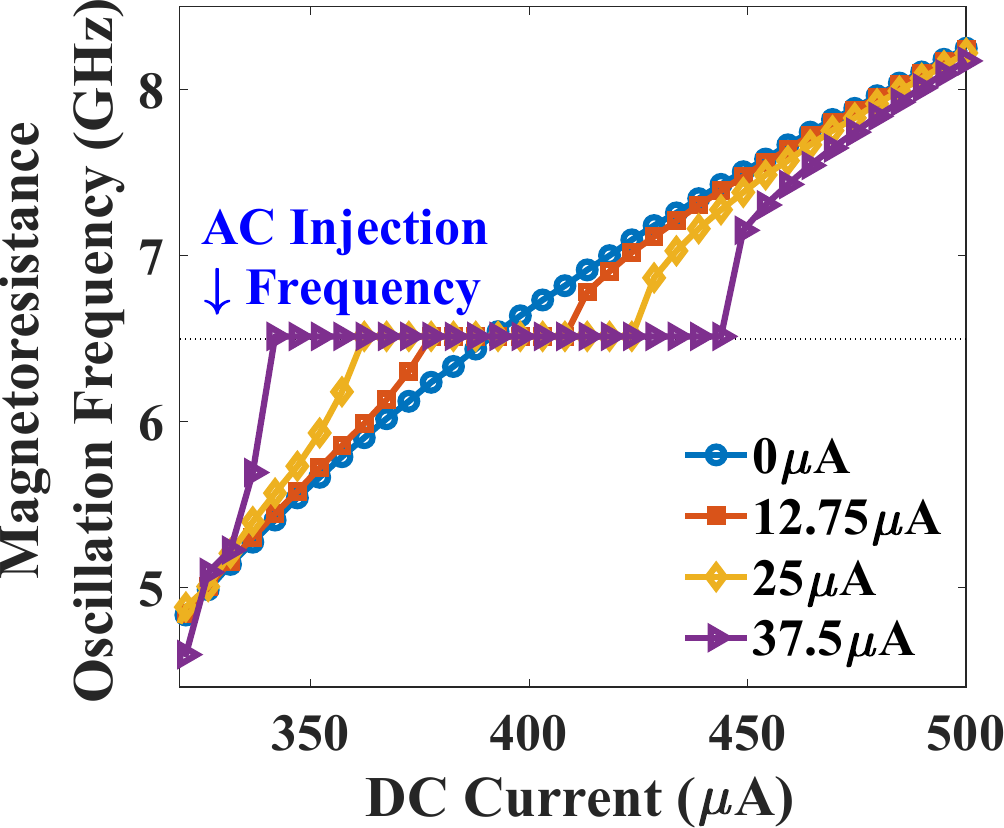}\label{fig:freq_vs_dc_Current}}\quad 
    \subfigure[]{\includegraphics[width=0.5\columnwidth]{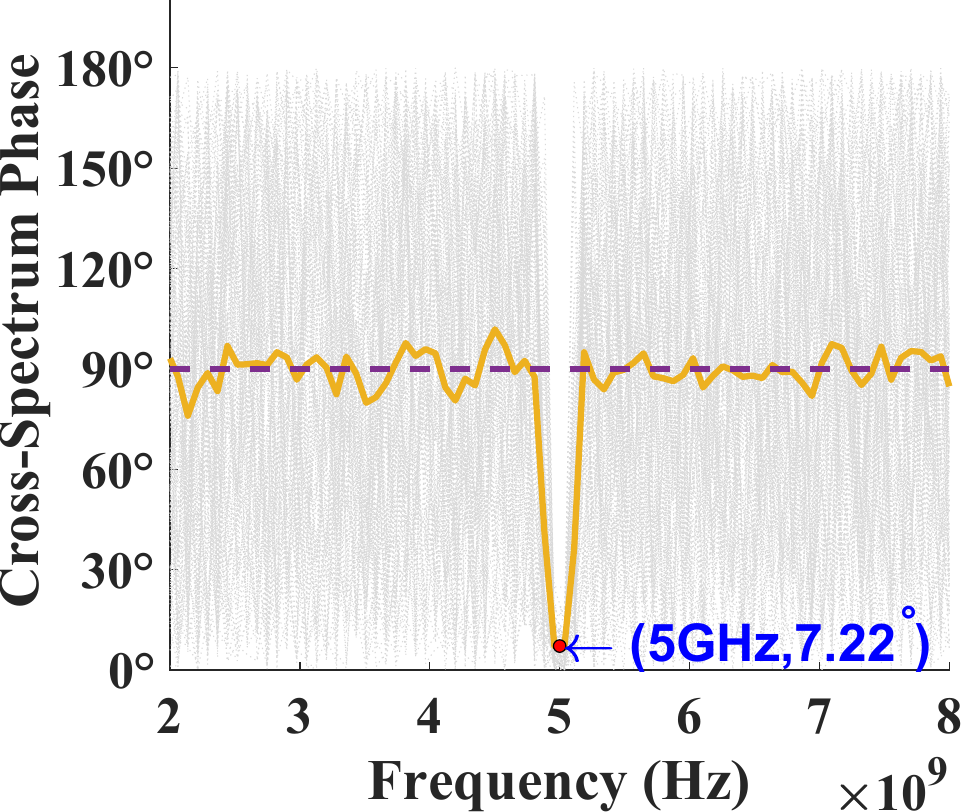}\label{fig:100 simulation data}}
     \subfigure[]{\includegraphics[width=0.5\columnwidth]{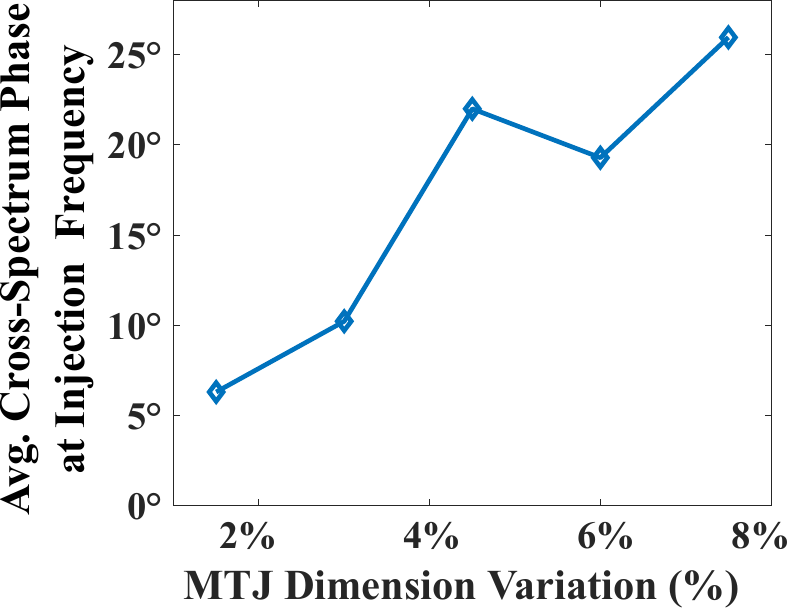}\label{fig:phase_corr_mu_and_Sigma}}
\caption{(a) Oscillator frequency plotted against the DC current input to the device. Higher AC amplitudes lead to increased DC locking range at the injected RF signal of $6.5GHz$ frequency. (b) Cross-spectrum phase for $100$ independent stochastic LLGS simulations of two noisy MTJ neurons, under RF injection of $5 GHz$. Average CPSD phase indicates tight phase-coupling at the required frequency with un-correlated activity at other frequencies. (c) Average cross-spectrum phase at the injection frequency accounting for device dimension variations.}
\end{figure*}

\subsection{Phase Synchronization of MTJ Oscillator Neurons}

The electrical analogue of Fig. 1(a) is shown in Fig. 3, where the MTJs represent the oscillatory neurons present in a particular network. The neurons share a HM layer which acts as the common substrate for the driving astrocyte signal. The current flowing through the HM has two components - a DC current input which determines the free-running frequency of the oscillator and a radio-frequency signal which represents the astrocyte input. Fig. 4(a) highlights the oscillation characteristics of the MTJ. The DC current controls the precession frequency in absence of other inputs. This DC input is analogous to the external stimulus determining the frequency of neuron oscillation in a particular network. In the absence of the RF signal, all the neurons oscillate at the same frequency (dependent on stimulus magnitude or DC current) but out-of-phase due to thermal noise. Upon the application of the  external RF astrocyte signal, the device oscillation locks in phase and frequency to this input. Higher peak-to-peak amplitude of the astrocyte locking signal increases the locking range of the device. It is worth mentioning here that the locking frequency of neurons in a particular network is dependent on the stimulus and astrocytes only induce phase locking. Therefore the alternating astrocyte signal flowing through the HM layer can be generated from a separate astrocyte device that is driven by the corresponding DC input of the network, thereby ensuring independent phase and frequency control. The astrocyte device is interfaced with a Reference MTJ and a voltage-to-current converter to drive the alternating current signal through the common HM layer. The Reference MTJ state is fixed to the AP state (by ensuring that the read supply voltage, $V_{DD} = 0.65V$ is not high enough to write the MTJ state) and forms a resistive divider with the oscillating Astrocyte MTJ resistance. Therefore, the gate voltage of the interfaced PMOS transistor, $V_G = \frac{R_A}{R_A+R_{REF}}V_{DD}$ where $R_A$ is the Astrocyte MTJ resistance and $R_{REF}$ is the Reference MTJ resistance, also varies accordingly, which in turn, modulates the current flowing through the common HM layer proportionally.

In order to evaluate the degree of phase synchronization in presence of thermal noise, we consider two MTJ devices lying on top of a common HM layer at room temperature. 
Cross-correlation metric is evaluated for the two MTJ output signals to measure the similarity among them as a function of displacement of one relative to the other. Considering two time-domain functions $x(t)$ and $y(t)$, whose power spectrum density (PSD) is given by $S_{xx}(\omega)$ and $S_{yy}(\omega)$ respectively, their cross-correlation is defined by:
\begin{equation} \label{eqn:Cross-correlation}
R_{xy}(\tau)=(x\star y)(\tau)=\int_{-\infty }^{\infty }\overline{x(t-\tau)}y(t)\, dt 
\end{equation}
where, $\overline{x(t)}$ represents the complex conjugate of $x(t)$ and $\tau$ denotes the lag parameter. Further, cross-power spectral density (CPSD) is defined as the Fourier transformation of cross-spectrum in~\eqref{eqn:Cross-correlation} and is given by:
\begin{equation} 
\label{eqn:CPSD}
S_{xy}(\omega) = \int_{-\infty }^{\infty } R_{xy}(t) e^{-j\omega t}\, dt
\end{equation}
$S_{xy}$ comprises of both magnitude and phase ($\angle$) information at different frequencies present in $\begin{bmatrix}
\omega
\end{bmatrix}$ vector. When two signals are phase synchronized, the cross-spectrum phase vector becomes zero, indicating high correlation.
Such a property is highlighted in Fig. 4(b) where $100$ independent stochastic-LLGS simulations are performed for two neuronal devices placed on a common HM layer with a $5 GHz$ injected RF current.  
Cross-spectrum phase at the injection frequency, i.e. $5 GHz$ converges close to zero. Average cross-spectrum phase is also shown in the plot depicting tight phase-coupling between the neurons at the injection frequency. Notably, a sharp reduction of average phase offset to just $7.22^{\circ}$ at $5 GHz$ is observed compared to $90^{\circ}$ for other frequencies, thereby establishing the robustness of the synchronization scheme. 
Additionally, the impact of non-idealities like device dimension variations on the phase coupling phenomena is evaluated in Fig. 4(c). The results are reported for 50 independent Monte-Carlo simulations with variation in both the length and width of the MTJ. Each Monte-Carlo simulation consisted of 50 stochastic LLGS simulation for the average cross-spectrum phase calculation. The phase correlation between the device oscillations remains reasonably high even with $7.5\%$ variation in both length and width dimensions of the MTJ. Related discussions on oscillator dynamics with respect to perturbative current and correspondence of the results with the Kuramoto model for oscillator synchronization is provided in the Supplementary section.


\begin{figure*}[ht!]
\center
\includegraphics[scale=0.76]{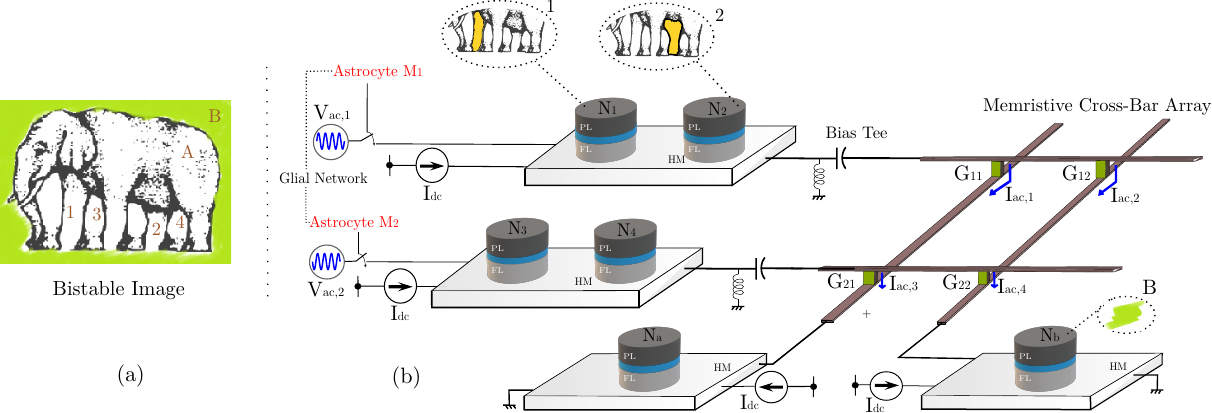}
\caption{\label{fig:Final_network} (a) The optical illusion induces confusion in the viewer concerning association among different apparent limbs with the body and the background (\textit{Courtesy of Roger Shepard's ``L'egsistential paradox"}) \cite{Elephant_image}. (b) MTJ system architecture depicting hierarchical organization of neurons. The illustrated binding problem is mapped to this hardware with one possible interpretation shown. The connection between different neuron layers is implemented by the memristive cross-bar array with initially untuned synaptic weights. Unsupervised STDP learning rule causes the weights to evolve, making the network to finally elicit synchronous responses post-training.   
}
\end{figure*} 
\section{\label{sec:Binding Problem} Binding problem }
\subsection{Problem Formulation}
Next, we discuss a renowned problem which is envisioned to be solved by neural synchronous activity. Amongst the most intriguing themes of neuro-psychological studies is the ``binding problem" (BP) \cite{Role_of_Glial_cells_in_learning_and_cognition, Binding_problem_discussed}. It concerns with how different attributes of sensory information are encoded, processed and perceived for decision-making by the human brain circuits. With a now widely accepted viewpoint of distributive computing and segregated processing for different features (especially visual) and later integration into a unified percept via re-entrant connections \cite{Binding_Segregation_1,Binding_Segregation_2}, we have progressed further towards understanding cognition. 
Primate brains have evolved to continuously assimilate the voluminous perceptive information available in their social setting and find a best fit for the primate's goals in the quickest manner. This training and growth, although very crucial in most situations -- sometimes also leads to ``misbinding" \cite{Misbinding}. In particular, optical illusions, such as shown in Fig.~\ref{fig:Final_network}(a), exploit the feature patterns ingrained in the human visual percept, causing misbinding. The figure is a bistable portrait of an elephant, or an overlap of two (seemingly) possible interpretations, obtained by associating different body parts to other features of the image. For instance, the labels 1 and 2 can be viewed associated with the body (A), while 3 and 4 to the background (B) to paint one such possible interpretation. The other interpretation can be visualized if the roles A and B are reversed. For an in-depth discussion, interested readers are directed to Ref. \cite{brenden_thesis, Memristor_optical_illusion}. In this work, we do not address the clustering mechanism of labels 1-2 and 3-4. This labelling and identification can be potentially attributed to the agent's visual attention. In particular, attention captures the most relevant information present in a space-time lapse by masking (filtering) off the distractor areas, while performing feature labelling of the cropped scene \cite{Adam_Attention2017_HART}. 
Assuming that attention performs the role of spatio-temporal integration among such multiple attributes captured by a visual scene, synchronous activity in the neurons is considered as the underlying mechanism in brain to create a coherent episode of perception, and perhaps cognition. Indeed, it is now becoming more evident that cognitive processes like attention and behavioral efficiency elicit targeted synchronous activity in different brain regions tuned to responding towards different spatial and featural attributes of the attended sensory input \cite{attention_and_sync_activity1, attention_and_sync_activity2}. 

\subsection{Hardware Mapping}

In order to correlate our spin-orbit torque oscillator phase synchronization due to astrocyte injection locking in the context of ``temporal binding", we consider a network as shown in Fig. \ref{fig:Final_network}(b). 
Adhering to the currently prominent view of hierarchical organization in the neural assemblies, spin-torque neurons $N_1, N_2, N_3, N_4$ here are dedicated to processing simple attributes, while $N_a$ and $N_b$ after receiving inputs from previous layers perform complex feature processing corresponding to the assigned task. In reference to potential processing applications like cognitive feature binding, each spin-orbit torque neuron in the network represents the corresponding feature in the elephant's bistable image, previously shown in Fig. \ref{fig:Final_network}(a). 
All neuronal devices are mounted atop a HM with $I\textsubscript{dc}=420 \mu A$ DC drive ($f\textsubscript{free}=7.05 GHz$). The network utilizes two different injection signals with the same frequency of $7.05GHz$ with $180^o$ phase difference (corresponding to the two different interpretations/configurations of the bistable image). Here, we use two RF voltage sources, namely $V\textsubscript{ac1}$ and $V\textsubscript{ac2}$ with amplitude of $250mV$. The connection between the two neuron layers is achieved by means of a resistive synaptic cross-bar array. We combine the concepts of bio-inspired unsupervised Spike-Timing Dependent Plasticity (STDP) \cite{bi1998synaptic} and astrocyte induced neural phase synchrony to automatically enable the network to learn to elicit such behavioral patterns, on the fly. The developed system sets off from an unlearnt state where all neurons have an independent response and remain unsynchronized in phase. However, upon system activation (and consequently astrocyte RF injection), the architecture eventually learns to bind the different possible configurations for the visual scene through phase correlation to either $V\textsubscript{ac1}$ or $V\textsubscript{ac2}$. It is to be noted that neurons $N\textsubscript{1}, N\textsubscript{2}, N\textsubscript{3}, N\textsubscript{4}$ comprise of pre-neurons while $N\textsubscript{a}$ and $N\textsubscript{b}$ are post neurons, separated by the resistive cross-bar array. Ultimately, a tight phase and frequency locking is observed among a particular pair of pre-neurons ($N_1, N_2$ and $N_3, N_4$) and post-neurons ($N_a$ and $N_b$). Due to random thermal fluctuations, the devices can converge to either of the two possible configurations for the bistable image, thereby illustrating the concept of optical illusion. The work can potentially pave the way for efficient hardware realization of coupled neuron-synapse-astrocyte networks enabled by compact neuromimetic devices.


\subsection{Learning Phase Correlation}

The premise for triggering the synchronous activity via astrocyte is accredited to the sensory attention as discussed before, and can be mapped in our proposed system to the amplitude of RF injection signal. Similar to better binding observed with increased attention, larger amplitudes lead to improved neural coupling. The strength of each input current to $N_{a}$ and $N_{b}$ is controlled by the synaptic conductances $G\textsubscript{11} - G\textsubscript{22}$ of the memristive cross-bar array as shown in Fig.~\ref{fig:Final_network}(b). Implementation of such cross-bar arrays with in-situ STDP learning has been previously explored for spintronic devices \cite{sengupta2016hybrid,sengupta2017encoding} and other post-CMOS technologies \cite{kuzum2011nanoelectronic,jo2010nanoscale,saha2021intrinsic}. It is worth mentioning here that each cross-connection also features a prior filtering ``bias tee" to eliminate any possible DC current interactions among different devices. The DC paths of  the bias tee are terminated to ground, while the AC signals get passed on to the cross-bar for coupling. Elaborating, the input AC current to the $j_{th}$ post-neuronal device (considering HM resistance to be considerably lower in comparison to the synaptic resistances at each cross-point) can be described by Eq. (\ref{eqn:AC_current_jth_neuron }) as: 

\begin{equation}
\label{eqn:AC_current_jth_neuron }
I_{ac,N_j}(t)= \sum_{i} G_{ij}.V_{i}(t)
\end{equation}

We now elucidate how our proposed architecture captures the essence of the optical illusion problem, shown in Fig. \ref{fig:Final_network}, in reference frame of an observer. Specifically, the system should be able to  adapt and converge to one of the possible interpretation discussed above.
In particular, biologically inspired unsupervised STDP principles are used to train the programmable synaptic conductances ($G\textsubscript{11} - G\textsubscript{22}$) in the cross-bar architecture for this purpose. The STDP weight (conductance) update equations are given by: $\Delta w = \eta_{+}w$ exp($\frac{-\Delta t}{\tau_{+}}$) (for $\Delta t > 0$) and  $\Delta w = \eta_{+}w$ exp($\frac{\Delta t}{\tau_{+}}$) (for $\Delta t < 0$), where $\eta_{+}$ and $\tau_{+}$ are learning hyperparameters, $\Delta w$ is the synaptic weight update and $\Delta t$ is the timing difference between the spikes corresponding to the selected post- and pre-neuron. The positive learning window ($\Delta t > 0$) update occurs whenever a post-neuron fires while the negative learning window ($\Delta t < 0$) update occurs at a pre-neuron firing event. It is worth pointing out here that we use a symmetric STDP learning rule in this work, i.e. the synaptic weight is potentiated for both the positive and negative learning windows. This is in contrast to the more popular asymmetric STDP observed in glutamatergic synapses \cite{bi1998synaptic}, typically used in neuromorphic algorithms \cite{diehl2015unsupervised}. While symmetric STDP has also been observed in GABAergic synapses \cite{woodin2003coincident}, further neuroscience insights are required to substantiate the exact underlying mechanisms and cause of this plasticity. Asymmetric STDP is useful in application domains requiring temporal ordering of spikes, i.e. a pre-synaptic neuron spike will trigger a post-neuron spike. However, for our scenario, a temporal correlation is crucial irrespective of the sequence, which is enabled by the symmetric STDP behavior. Implementation of symmetric STDP in memristive cross-bar arrays can be easily achieved by proper waveform engineering of the programming voltage applied across the synapses \cite{sengupta2016hybrid,serrano2013stdp}. The cross-bar resistances are considered to have an ON/OFF resistance ratio of $10$. The different input spike trains are derived from each device's magnetoresistance (MR) where a spike is triggered when the MR crosses its mean-value of 2$K\Omega$.  Because $N\textsubscript{1}$ ($N\textsubscript{3}$) and $N\textsubscript{2}$ ($N\textsubscript{4}$) share a common HM, either of them can be used to extract the pre-neuron spikes during the weight update period. Besides STDP, a lateral inhibition effect \cite{diehl2015unsupervised} is utilized. Whenever a spike occurs for any pre-neuron (post-neuron), the corresponding row (column) weights of the array are potentiated. However, the remaining rows (columns) are depressed proportionately. The lateral inhibition weight update equations are given by: $\Delta w = - \eta_{-}w$ exp($\frac{-\Delta t}{\tau_{-}}$) (for $\Delta t > 0$) and  $\Delta w = - \eta_{-}w$ exp($\frac{\Delta t}{\tau_{-}}$) (for $\Delta t < 0$), where $\eta_{-}$ and $\tau_{-}$ are learning hyperparameters, $\Delta w$ is the synaptic weight update and $\Delta t$ is the timing difference corresponding to the symmetric STDP weight update for the row or column which experiences weight potentiation. The lateral inhibition scheme is a simple extension of the synaptic programming voltage waveform engineering used in prior work \cite{serrano2013stdp,indiveri2011neuromorphic,sengupta2016hybrid}. 
During the learning phase, this lateral inhibition effect causes the neuron under study to start responding selectively towards a specific configuration. This, in turn, enables the network to later converge to one of the interpretations for Fig. \ref{fig:Final_network}(a), as mentioned previously. The network simulation parameters are outlined in Table II. The tabulated time-constants are measured with respect to the time-step for LLG simulation.
\subsection{Simulation Results}

\begin{table}[t!]
\center
\centerline{Table II: Learning Simulation Parameters}
\vspace{2mm}
\begin{tabular}{c c}
\hline \hline
\bfseries Parameters & \bfseries Value\\
\hline
Time-step for LLG simulation & $0.1ps$\\
STDP learning rate, $\eta_+$ & $0.25$ \\ 
STDP time constant, $\tau_+$  & $5$ \\
Inhibition learning rate, $\eta_{-}$ & $0.15$\\
Inhibition time constant,  $\tau_{-}$ & $5$\\
Maximum synapse resistance in cross-bar array & $25k\Omega$ \\ 

\hline \hline
\end{tabular}\\ 
\label{tab:network_params}
\end{table}

\begin{figure}[t]
\center
  \subfigure[]{\includegraphics[width=0.61\columnwidth]{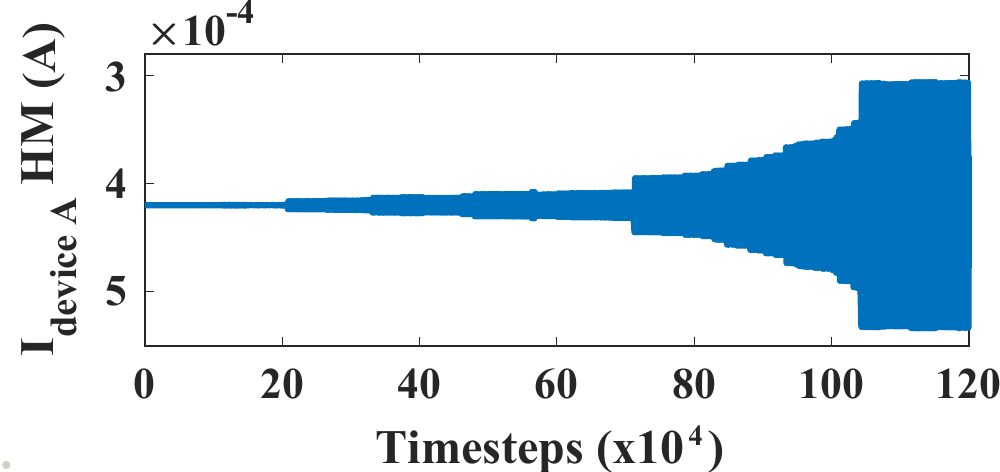}\label{fig:I_DEVICEa}}
    \subfigure[]{\includegraphics[width=0.6\columnwidth]{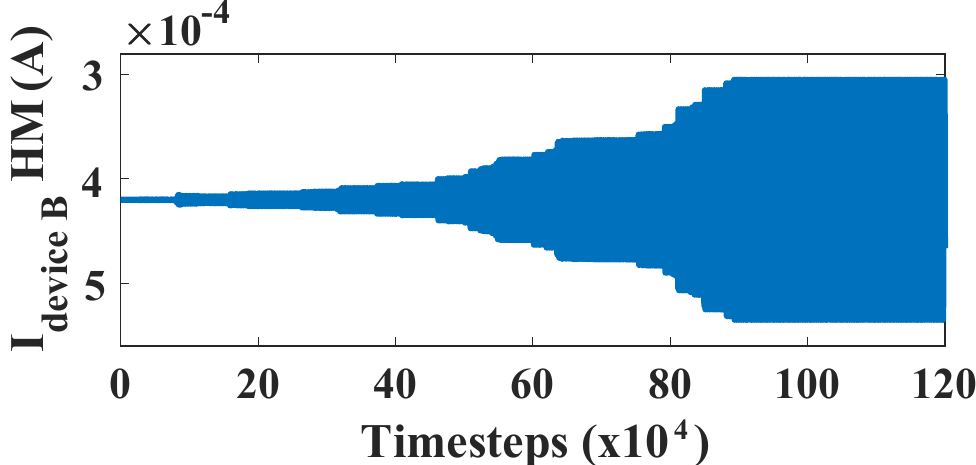}\label{fig:I_deviceB}}
    \subfigure[]{\includegraphics[width=0.58\columnwidth]{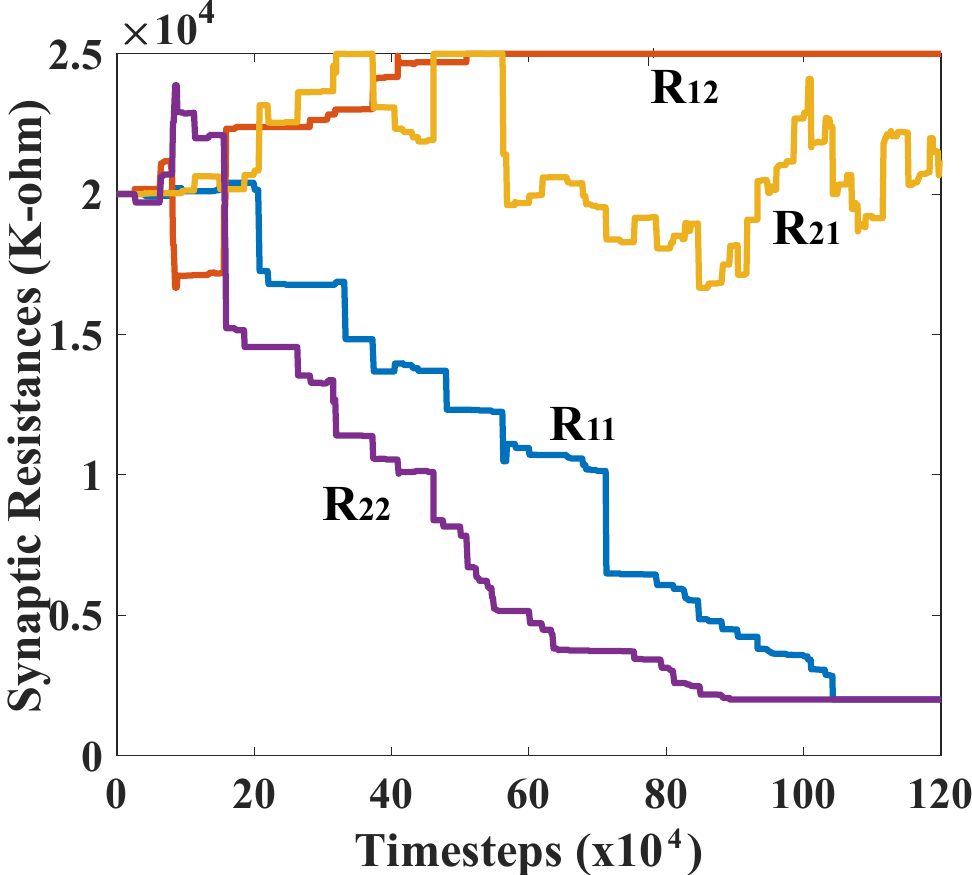}\label{fig:Weights}}

\caption{(a-b) The temporal evolution profiles of the net currents (DC+AC)  flowing through the heavy metal for devices A and B are shown. The increasing AC amplitudes about the mean DC value can be seen. The relatively flattened envelopes post-learning suggest that the post-neuron devices are dominantly locked to one of the frequencies. (c) Temporal evolution of the cross-bar resistances during the learning process is shown.}        
\end{figure}


The net currents for devices A and B, evolving through time, is portrayed for one of the simulations in Figs. 6(a) and 6(b) respectively.
Meanwhile, the corresponding synaptic resistances for the network are plotted in Fig. 6(c) to elucidate the learning process discussed previously. 
The learning phase for the simulation is plotted as a function of timestep of the LLG simulation of the MTJ devices ($0.1ps$). Observing the temporal profiles, an interesting deduction can be formulated, confirming that the different post-neurons get dominantly locked to  different injection frequencies. The two sinusoids, being initially out of phase and adding up in comparable amounts for post-neurons, result in very low net currents. But, as the learning progresses, it becomes clear that one of the frequency gets dominant for a particular post-neuron, and thus the envelope tends to flatten in the end. It is worth mentioning here that the synaptic learning simulation in this work was performed from an algorithmic standpoint in a technology agnostic fashion. Depending on the underlying synapse technology, prior proposals for peripheral design for STDP learning needs to be considered \cite{sengupta2016hybrid,serrano2013stdp}. Since the focus of this article is on the MTJ neural synchrony aspect, we did not consider any specific synaptic device programming delay constraint (which is reflected in the instantaneous state changes of the synaptic connection strengths in Fig. 6(c)). In reality, from a system design perspective, we need to have interleaved synaptic device state update phases that do not interfere with the neuron oscillation behavior (for instance, through decoupled write-read phases of three-terminal synaptic devices \cite{sengupta2016hybrid}). The convergence was also not affected with reduced programming resolution of the synaptic connections (4-bits), thereby indicating resiliency to quantization \cite{hu2021quantized}.


\begin{figure}[t!]
\center
  \subfigure[]{\includegraphics[width=0.49\columnwidth]{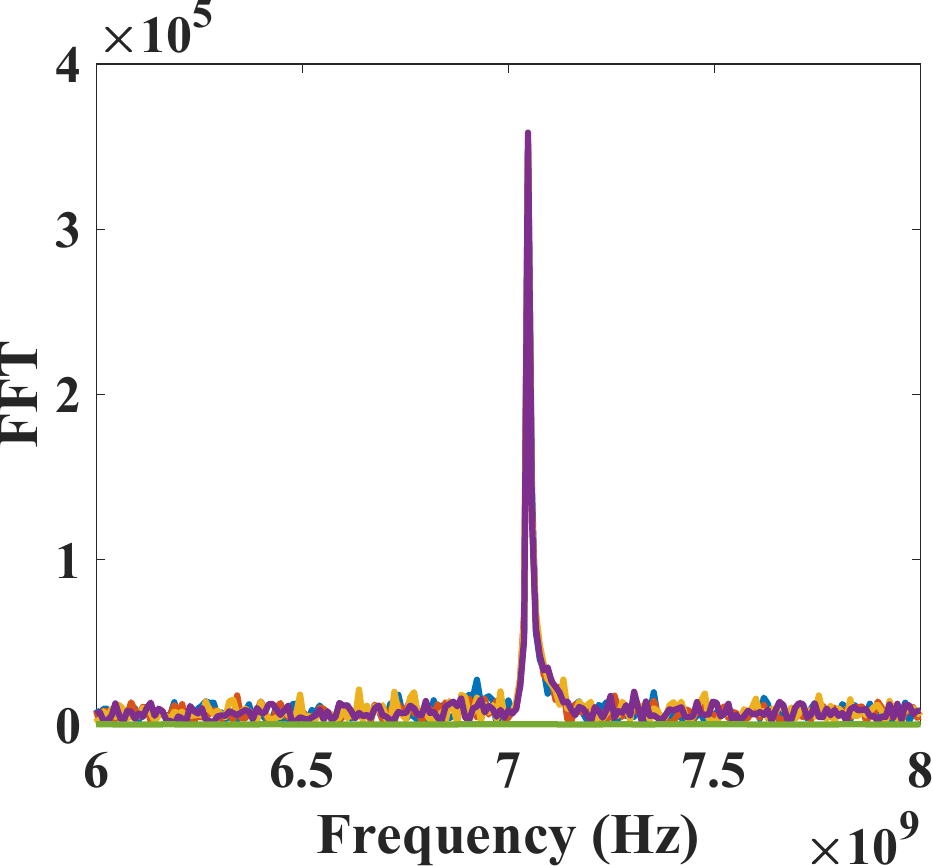}\label{fig:FFT_network_config1 }} 
    \subfigure[]{\includegraphics[width=0.49\columnwidth]{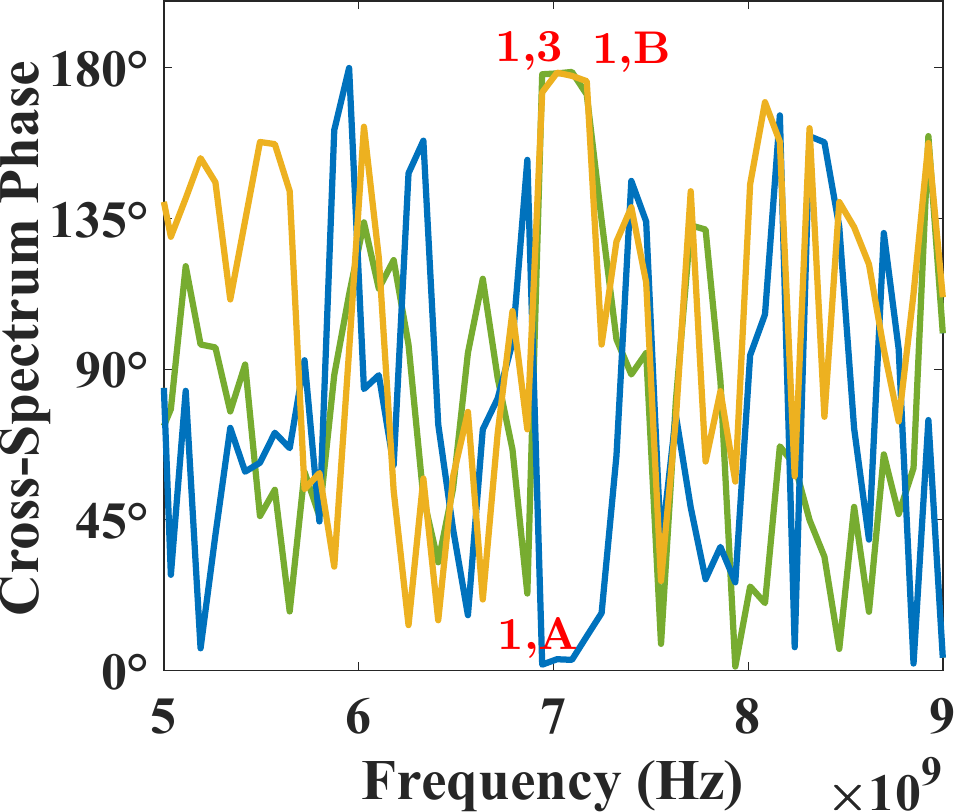}\label{fig:FFT_network_config2}}
\caption{(a) FFT plots for all devices for one of the two possible configurations are shown post-learning. (b) Cross-spectrum phase for devices-pairs 1-3 ($178.53^{\circ}$), 1-A ($3.35^{\circ}$) and 1-B ($178.3^{\circ}$) are plotted to show the phase-locking nature of the network post-learning at the injection frequency of $7.05 GHz$.  
}
\end{figure}

\begin{figure*}[t!]
\center
\includegraphics[scale=0.92
]{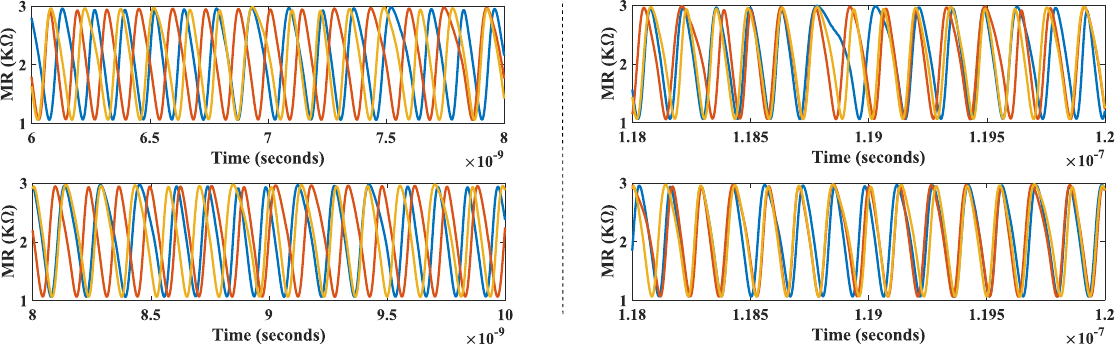}
\caption{\label{fig:Magneto-resistance profile} Temporal profile for the devices in the network (shown in Fig. \ref{fig:Final_network}(b)) before (left) and after synchronization (right) are depicted for one particular configuration. Astrocyte functionality activates the synchronous regime causing learning to occur and subsequently coherent neural patterns are achieved for this configuration (a stochastic event). Devices $N_1$, $N_2$ and  $N_a$ (top-right panel) lock to injection signal with $\phi=0^{\circ}$, while devices $N_3$, $N_4$ and $N_b$ reveal concerted neural patterns in conjunction to $\phi =180^{\circ}$ injection signal. (bottom-right panel).   }
\end{figure*} 

Cohesing to one of the percept should surmise of a random event to provide equal chance for any of the two possible configurations to develop. Indeed, it is observed in our network that the synchronization occurs for random first and second layer neurons, post-training. Such a phenomenon can be accredited to the natural thermal fluctuations in our system, which tend to perturb the MTJ device's periodic nature. Fig. 7(a) and 7(b) respectively depict the FFTs and cross-spectrum phase for various devices in the network for one such possible configuration upon learning termination. Specifically, cross-spectrum phases for device-pairs 1 \& A (blue curve), 1 \& 3 (yellow curve) and 1 \& B (green curve) in Fig. 7(b) are plotted to highlight that device 1, 2 and A get locked in phase at the injection frequency ($7.05GHz$) while being completely out of phase with devices 3, 4 and B for the considered configuration.  


Fig.~\ref{fig:Magneto-resistance profile} plots the temporal profile of device magnetoresistance (MR) for $N_1, N_2$ and $N_a$ devices in the top panel, along with MR of $N_3$, $ N_4$ and $N_b$ devices shown in the bottom panel.
Initially all neuronal devices, albeit operating at the same free-running frequency ($f\textsubscript{free}= 7.05GHz$), elicit un-correlated phases, and hence temporal spike response due to devices' inherent thermal noise.
After the astrocyte AC signal injection and STDP learning commences, it is observed that the devices $N_1$ ($N_3$) and $N_2$ ($N_4$) achieve a gradual coherent phase along with device $N_a$ ($N_b$), getting locked to the respective injection signal, as can be clearly seen in the right panels. The subsequent cross-correlation phase at the $7.05 GHz$ injection frequency post-synchronization averages to $1.6232^\circ$ for the three-possible temporal profile pairs among $N_1, N_2$ and $N_a$ ($N_1\star N_2$: $ 0.88^\circ, N_2\star N_a$: $ 2.136^\circ$, and $N_1 \star N_a$: $1.856^\circ $). Likewise, $N_3, N_4$ and $N_b$ after learning, achieve an average cross-phase of $1.848^\circ$. Bio-physically equivalent, this can be interpreted as a tight correlation among the attributes $1,\,2$ and $\,A$, corresponding to one of the interpretations of the bistable image. Finally, an increasing phase-mismatch is visible in neuronal outputs of all devices if the synchronization is revoked by the astrocyte, and the devices revert to their uncorrelated original free running frequency. This can be attributed to a diverted attention towards the sensory modal-input features leading to the impairment in correlated activity.

\section{\label{sec:Discussion} Discussion}

Even though this work proves to be a good preliminary framework for emulating such brain-like functions, more investigation is required for decoding the neural code in such processes along with integrating these insights in Artificial Intelligence (AI) systems. For instance, selectivity bias towards some features among the myriad available sensory information, and, reductionism (down-streaming) of such higher-level modal inputs to local neuronal groups in the hierarchical structure, is poorly understood. There have been some efforts to study such processes using a reverse approach, where robots like Darwin VIII, inspired by the re-entrant neuroanatomy and synaptic plasticity, are developed and trained on visual mode data \cite{Darwin_robot}. In agreement with our work, they show synchronous activity binds different representative features of the detected object. Incorporating such connections in our system can be explored to further bridge the gap between real cortical networks and the respective inspired models. 
Supported by both neuroscience research and AI hardware developments, coupled astrocyte-neuron network architectures can potentially pave the way for a new generation of artificial cognitive-intelligence.

\section*{Data Availability Statement}

The original contributions presented in the study are included in the article, further inquiries can be directed to the corresponding author/s.

\section*{Author Contributions}

All authors contributed equally to the writing of the paper, developing the concepts and performing the simulations.

\section*{Supplementary Information}

In order to better understand and substantiate the phase synchrony of the spin-torque oscillators, we considered the phase resetting curve (PRC) of the devices. The PRC plots the relation between phase shift and the timing of perturbation that causes this phase shift and describes the phase sensitivity of nonlinear oscillators to external perturbations \cite{prcurve}. There are two types of PRCs. Type-I PRC is non-negative or non-positive curve while type-II PRC has both negative and positive components in the curve \cite{prcurve}.

\begin{figure*}[t]
\center
\subfigure[]{\includegraphics[width=0.6\columnwidth]{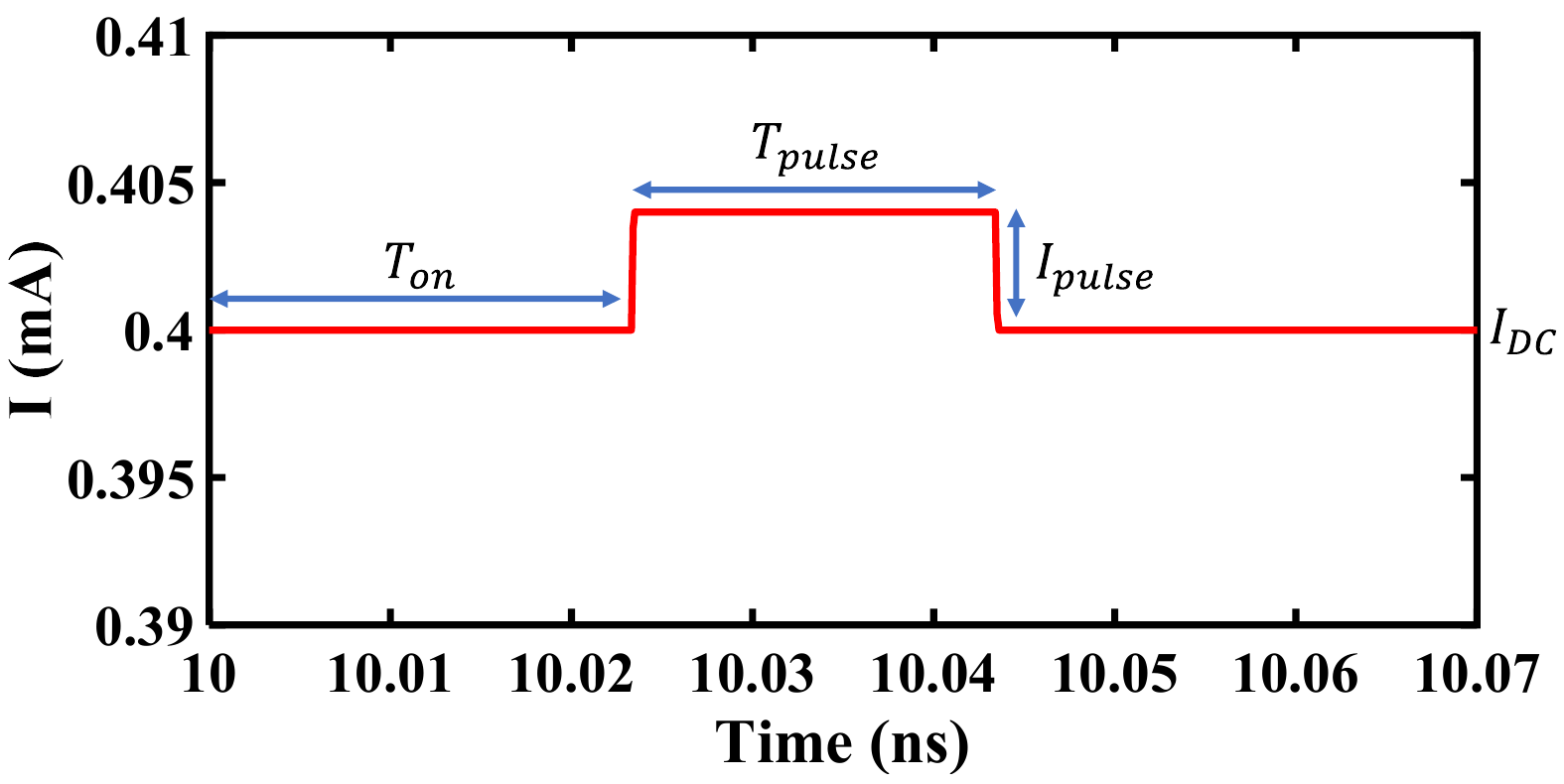}}\label{fig:Currentpulse}
\subfigure[]{\includegraphics[width=0.6\columnwidth]{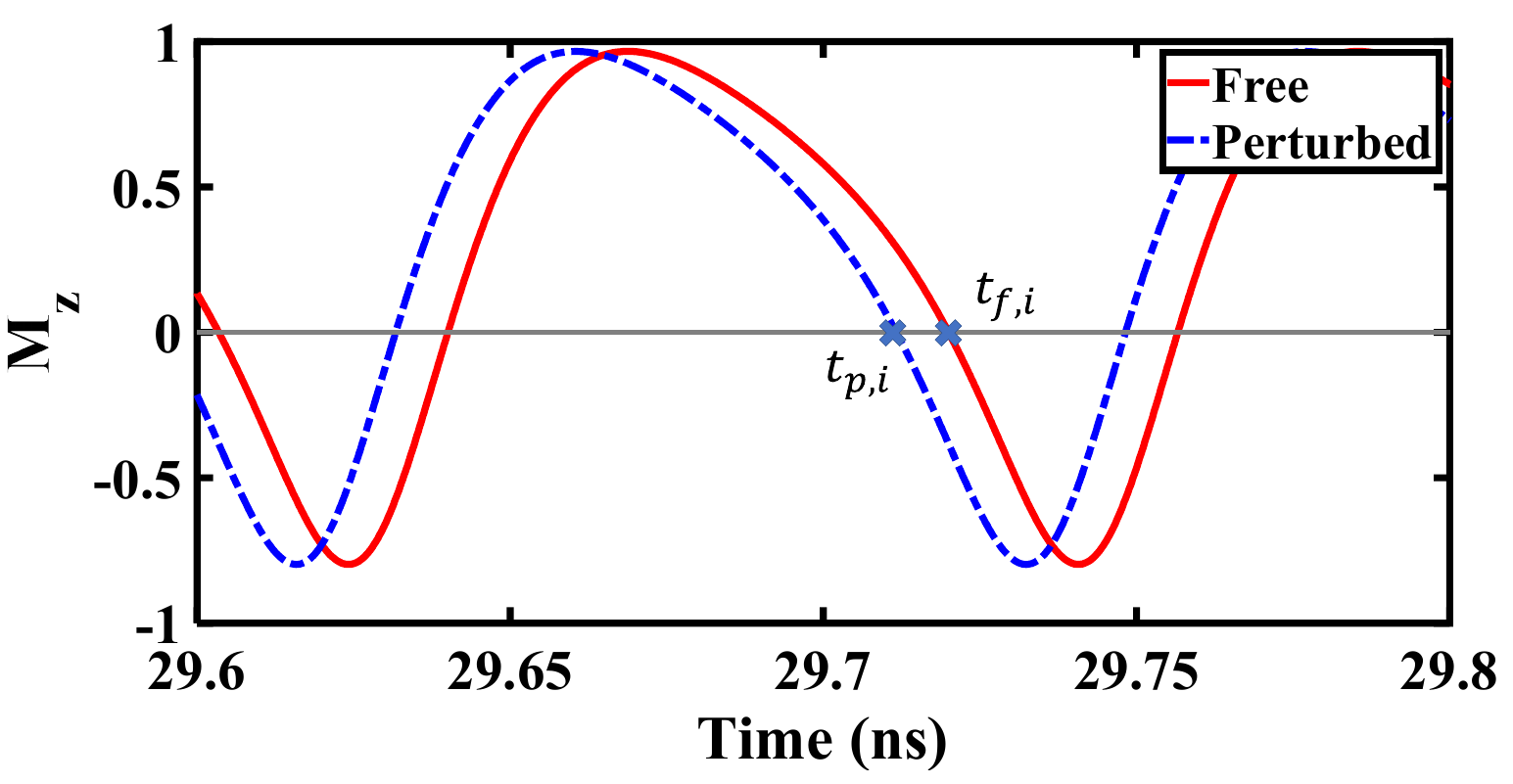}}\label{fig:CompareFP}
\subfigure[]{\includegraphics[width=0.6\columnwidth]{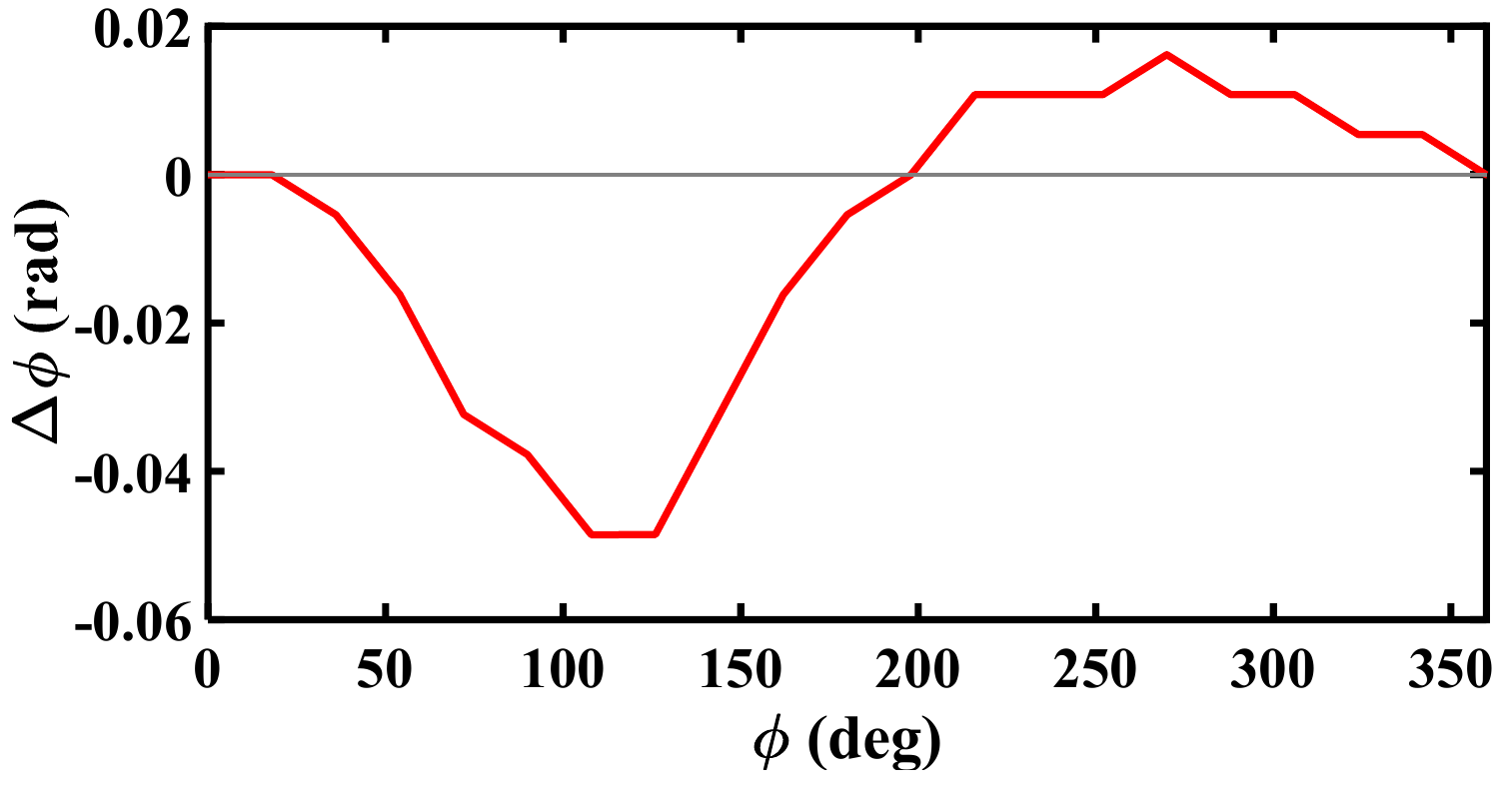}}\label{fig:Phasesens}
\caption{(a) Current pulse applied to the perturbed oscillator. $T_{on}$ is the time instant when the current pulse is applied, which disregards the first 10 ns (since the oscillator reaches stable oscillation in 10 ns). $T_{period}$ is the pulse width and $I_{pulse}$ is the current amplitude. (b) Magnetization component in easy axis of the free and perturbed oscillators. The $i$-th time-instant when the free (perturbed) oscillator's magnetization reaches 0 is noted as $t_{f,i}$ ($t_{p,i}$). Phase shift $\Delta\phi_i$ is determined by $2\pi (t_{f,i}-t_{p,i})/T_{period}$, where $T_{period}$ is the oscillation period. (c) PRC of the oscillator at $I_{DC}=400 \mu A$. Pulse amplitude is $I_{pulse} = 4\mu A$ and pulse width is $T_{pulse}=0.02 ns$.}
\end{figure*}

To obtain the PRC through numerical calculations, we compared two oscillators. One of the oscillators was free-running with only a DC current pulse while the other perturbed oscillator was subjected to a pulse current applied in addition to the DC current, as is shown in Fig. 9(a). The phase shift due to the perturbation, $\Delta\phi$, is obtained from $\Delta\phi_i = 2\pi\Delta t_i/T_{period}$, where $T_{period}$ is the oscillation period, $\Delta t_i = t_{f,i} - t_{p,i}$ is the $i$-th phase shift during the period (as shown in Fig. 9(b)). $\Delta\phi_i$ converges to a constant value after a period of time, which gives the value of $\Delta\phi$. $\phi$ is the phase of oscillator when perturbation is applied. The relation between $\Delta\phi$ and $\phi$ is the PRC of oscillator, which is plotted in Fig. 9(c). For our simulations, we considered $I_{DC} = 400 \mu A$ and $I_{pulse} = 4 \mu A$. Pulse width, $T_{pulse}$, is taken to be $20 \%$ of the oscillator period. The PRC shown in the figure is a type-II PRC. Oscillators with type-II PRC can be synchronized more efficiently under common noisy input compared to oscillators with type-I PRC, as is indicated in prior work \cite{PhysRevE.80.011911}.

To further understand the synchronization phenomena of a network of oscillators, we considered the Kuramoto model \cite{acebron2005kuramoto}. Coupling phenomena of spintronic oscillators for different schemes have been mapped to the Kuramoto model in prior works \cite{Flovik2016,Kabir2015}. A phase sensitivity function augmented Kuramoto modelling approach has also been adopted to explain synchronization of spintronic oscillators under the application of pulse currents \cite{nakada2016pulse}. The Kuramoto model dynamics is described by,
\begin{equation}
\dot{\theta_{i}} = \omega_{i} + \sum_{j\neq i}K_{ij}\sin{(\theta_{j}-\theta_{i})}
\label{eqn:kuramoto Equation}
\end{equation}
where, $\dot{\theta_{i}}$ is the actual frequency of the $i$-th oscillator, $\omega_{i}$ is the intrinsic frequency of the $i$-th oscillator, $K_{ij}$ is the coupling strength between the $i$-th and $j$-th oscillator, and $\theta_{i}$ is the phase of the $i$-th oscillator.

In the coupling scheme shown in Fig. \ref{fig:Multiple_Devices}, multiple neuron devices are mounted upon a heavy metal layer through which an AC signal is driven by an astrocyte device. This AC signal provides the coupling signal between the driver device and each target neuron device. On the other hand, there is no signal from the neuron devices to the driver astrocyte device or between the target neuron devices (due to negligible dipolar coupling). Denoting the driver device by index $0$ and the target neuron devices as $i > 0$, the coupling strength $K_{ij}$ for the $i$-th target oscillator is given by,
\begin{equation}
K_{ij} = \left\{
\begin{array}{rcl}
0 & & {j > 0}\\
f(I_{AC}) & & {j = 0}
\end{array} \right.
\label{eqn:coupling condition}
\end{equation}
where, $I_{AC}$ is the amplitude of the AC signal. The DC bias current, $I_{DC}$ of the oscillators control the intrinsic frequency $\omega_i$. In this case, for the driver astrocyte device, the actual frequency $\dot{\theta_0} = \omega_{0}$, and for the target neuron devices, $\dot{\theta_i} = \omega_{i} +K_{i0}\sin{(\theta_{0}-\theta_{i})}$, where $i > 0$. We consider a particular simulation setup of the Kuramoto model assuming a linear $\omega-I_{DC}$ relationship of the oscillators, similar to the one shown in Fig. 4(a). As shown in Fig. 10, the Kuramoto model oscillator coupling characteristics bear close resemblance to the spintronic oscillator coupling characteristics in Fig. 4(a), where the target oscillator actual frequency locks in to the driver frequency for a non-zero coupling strength.
\begin{figure}[t]
\center
\includegraphics[width=0.7\columnwidth]{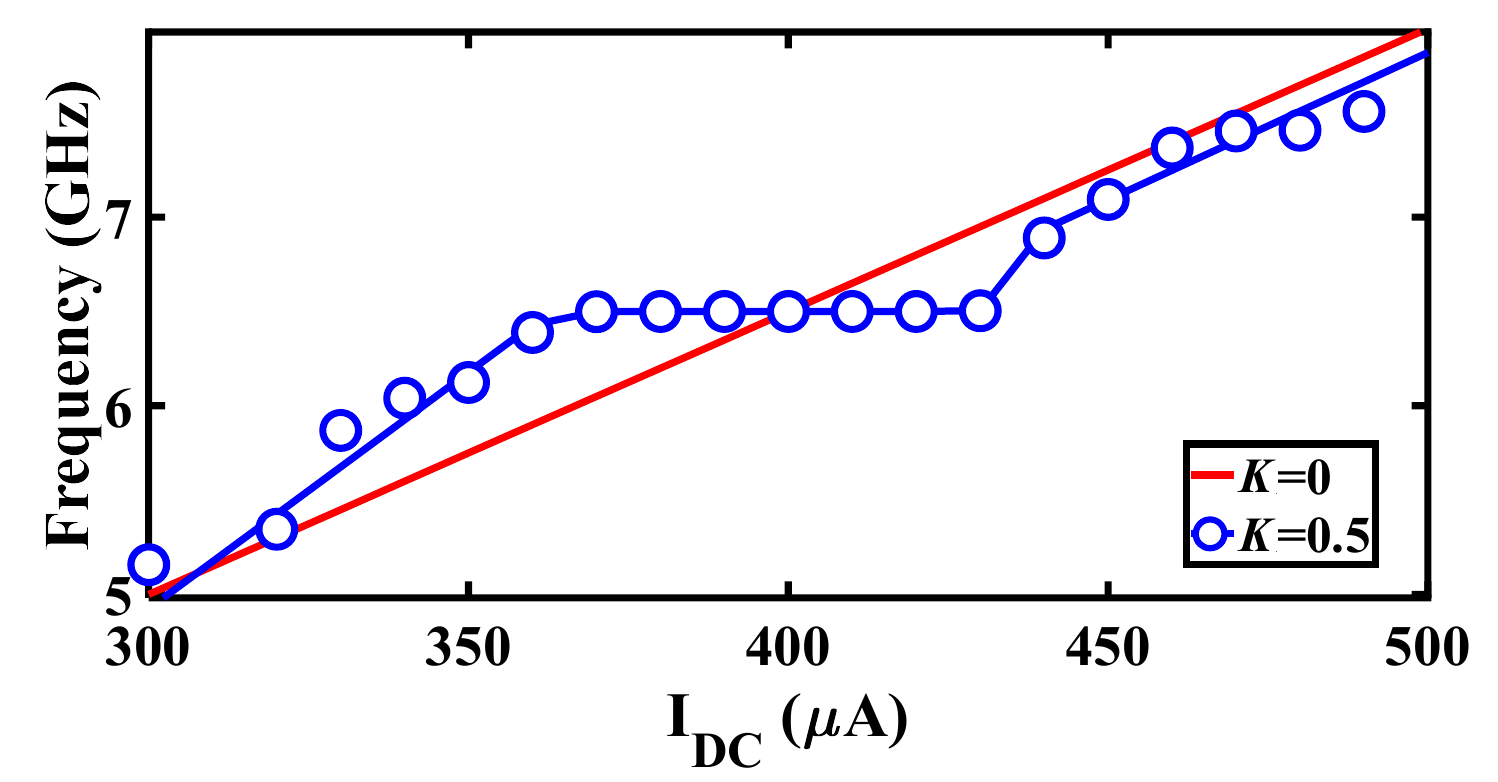}
\caption{\label{fig:FreSweep} Kuramoto model oscillator coupling characteristics for non-zero coupling strength closely resembles the spin-oscillator synchronization characteristics. The AC injection frequency is $6.5GHz$.}
\end{figure}

\section*{Acknowledgments}

The work was supported in part by the National Science Foundation grants BCS \#2031632, ECCS \#2028213 and CCF \#1955815.

\providecommand{\noopsort}[1]{}\providecommand{\singleletter}[1]{#1}%

\end{document}